# Observation of Large Enhancements of Charge Exchange Cross Sections with Neutron-Rich Carbon Isotopes

I. Tanihata[1,2]*, S. Terashima[1]*, R. Kanungo[3], F. Ameil[4], J. Atkinson[2], Y. Ayyad[2], D. Cortina-Gil[5], I. Dillmann[4], A. Estradé[3,4], A. Evdokimov[4], F. Farinon[4], H. Geissel[4], G. Guastalla[4], R. Janik[6], R. Knoebel[4], J. Kurcewicz[4], Yu. A. Litvinov[4], M. Marta[4], M. Mostazo[5], I. Mukha[4], C. Nociforo[4], H.J. Ong[2], S. Pietri[4], A. Prochazka[4], C. Scheidenberger[4], B. Sitar[6], P. Strmen[6], M. Takechi[4], J. Tanaka[2], H. Toki[2], J. Vargas[5], J. S. Winfield[4], H. Weick[4]

[1] *School of Physics and Nuclear Energy Engineering and IRCNPC, Beihang University, Beijing 100191, China*
[2] *RCNP, Osaka University, Ibaraki 567-0047, Japan*
[3] *Saint Mary's University, Halifax, NS B3H 3C3, Canada*
[4] *GSI Helmholtz Center, 64291 Darmstadt, Germany*
[5] *Universidad de Santiago de Compostela, Santiago de Compostela, Spain*
[6] *Comenius University, Bratislava, Slovakia*
*Email:* tanihata@rcnp.osaka-u.ac.jp

Production cross sections of nitrogen isotopes from high-energy (~950 MeV per nucleon) carbon isotopes on hydrogen and carbon targets have been measured for the first time for a wide range of isotopes ($A = 12$ to 19). The fragment separator FRS at GSI was used to deliver C-isotope beams. The cross sections of the production of N-isotopes were determined by charge measurements of forward going fragments. The cross sections show a rapid increase with the number of neutrons in the projectile. Since the production of nitrogen is mostly due to charge-exchange (Cex) reactions below the proton separation energies, the present data suggests a concentration of Gamow–Teller and/or Fermi transition strength at low excitation energies for neutron-rich carbon isotopes. It was also observed that the Cex cross sections were enhanced much more strongly for neutron-rich isotopes in the C-target data.

Subject index: D12, D13, D23, D27

Charge exchange (Cex) reactions such as (p,n) and ($^3$He,t) at intermediate and high energies bring about similar transitions as Fermi (F) and Gamow–Teller (GT) β decays, including transitions to higher excited states that could not be populated by β decays. Such transitions in nuclei near the stability lines have been studied and the building up of giant GT resonances have been discussed for mid-shell nuclei by Fujita et al. [1]. In that study, a change of transition strength was investigated by ($^3$He,t) reactions at 140 MeV/nucleon using even–even $T_Z= 1$ target nuclei in the $f_{7/2}$ shell. It was found that most of the transition strength was concentrated in the low-energy states of produced nuclei at the bottom of the shell, such as for the case of $^{42}$Sc. The transition to the isobaric analog state (IAS) and the first GT state carry most of the transition strength below 12 MeV. In contrast, the transition strength in high-energy excitations increases when the number of nucleons in the shell increases. For example, most of the strength is located from 6 to 12 MeV, i.e. the giant resonance region. The low-energy excited states including IAS contribute a very small amount to the total strength. This is due to a development of giant GT resonances at higher excitation energies. Fujita's experiment shows a gradual change of the strength distribution when the number of nucleons in a valence shell increases along the stability line. Interest remains in how the strength changes when only the number of neutrons increased to unstable nuclei.

Many studies have been reported, [2, 3] on the relationship between the β-decay strength and the cross section. These two observed values are related to each other and commonly parameterized as

$$\sigma = \hat{\sigma}_i F_i(q,\omega) B(i), \tag{1}$$

where $i$ = F or GT distinguishes the Fermi and GT transitions. The proportionality factor $\hat{\sigma}$ is called the unit cross section and may depend on the beam energy. The factor $F_i(q,\omega)$ describes the shape of the cross section distribution and goes to unity in the limit of zero momentum and energy transfer. We assume $F_i(q,\omega) \sim 1$ in the following discussion. The last factor, the beta-decay transition strength $B(i)$, is obtained from the beta-decay *ft* value. If the relationship between the beta-decay and (p,n) reactions is direct, the unit cross section is expected to be a slowly changing function of the mass of the nuclei, $A$. Taddeucci et al. [4] studied the unit cross section for many nuclei in a wide range of masses. Also, Sasano et al. [5] studied the unit cross section systematically for medium-heavy nuclei. Those papers reported that the unit cross section changes smoothly with mass number except for light nuclei. Taddeucci et al. observed a peculiar behavior in the strength for C isotopes. The unit cross sections for $^{12}$C, $^{13}$C, and $^{14}$C do not exhibit a smooth dependence on *A*, but vary greatly. This contrasts strongly with the fact that the same value of unit cross section can be used for transitions to states of different excitation energies within one nuclide. This peculiar behavior could be due to an uncertainty in the distorted wave impulse approximation used, though other possibilities cannot be rejected [4]. Therefore, studies with the longer chains of C-isotopes may shed light on the isotope dependence. The beam energy dependence of such relationships was studied by Fujiwara et al. by (p,n) and ($^3$He,t) reactions [6]. As an example, the ratio of the cross sections σ[$^{12}$C -> $^{12}$N(gs)]/ σ[$^{13}$C -> $^{13}$N(3.51)] was found to be constant for beam energies from 150 to 700 MeV/nucleon. It is generally observed that the unit cross section of a Fermi transition is much smaller (1/5 to 1/10) than that of GT transitions [4].

For understanding the r-process, i.e., the nucleosynthesis of the heaviest nuclei, the β-decay strengths of very neutron-rich nuclei are essential information. The total β-decay strength of very-neutron-rich nuclei is the sum of all transitions and is directly related to the transition strength that may be measured by Cex reactions using high-energy beams of neutron-rich nuclei. To date, no systematic measurements of Cex reactions have been made for neutron-rich unstable nuclei. The known smoothness of the unit cross section for heavy nuclei is advantageous for such a study.

Searches for IASs of very-neutron-rich nuclei have been reported for $^{11}$Li [7, 8] and $^{14}$Be [9] using (p,n) reactions and strong transitions to IASs have been observed for both these nuclei. However, studies have only been made of selected neutron halo nuclei and no systematic measurements have yet been reported. Charge exchange reactions with heavy ions and their relation to GT transitions have been discussed by Osterfeld et al. [10]. They discussed Cex reactions with particular reference to the ($^{12}$C,$^{12}$B) and ($^{12}$C,$^{12}$N) reactions under a strong absorption model. They found that the *L* = 0 transitions clearly reflect the strength of the GT transition. In contrast, a later theoretical study by Bertulani and Lotti [11] concluded that the determination of GT and Fermi strength from heavy-ion Cex is not necessarily straightforward. Consequently, systematic studies of Cex reactions with heavy ions, in addition to the (p,n) reaction are necessary, particularly for neutron-rich nuclei.

The present paper reports the first measurement of the isotope dependence of the production of nitrogen isotopes from high-energy incident carbon isotopes, $^{12-19}$C, on proton and carbon targets. Outgoing N isotopes were measured at near zero degrees, covering most of the scattering angles of projectile fragmentation. We call this cross section the charge-exchange reaction cross section ($σ_{ex}$) because the proton-transfer reaction cross section is expected to be much smaller than the cross section due to a charge exchange between a projectile and a target. This charge exchange at high energies in this study is expected to occur mainly though charged meson exchanges and proton–neutron exchange reactions [12]. Theoretical study was made for estimating the contribution from direct charge exchange due to central and tensor interactions and from sequential proton and neutron transfer [13]. Their calculations indicated that direct charge exchange is safely dominant for incident energies above 100*A* MeV. In the final states of the present experiment, only N isotopes were identified and thus the reaction is charge exchange restricted to produce N isotopes below the proton emission threshold. In contrast, neutron emissions from excited states do not change the Z value of the final state and thus those channels are included in the measured cross sections.

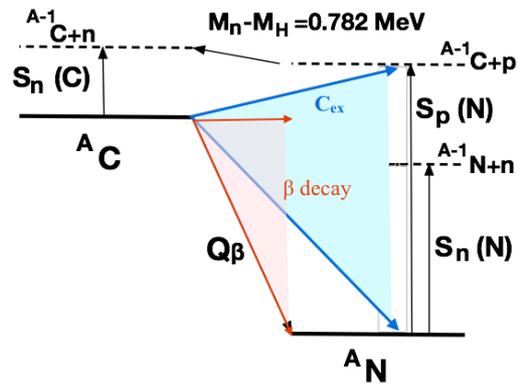

Fig. 1 Related nuclei and the relation between the separation energies ($S_n$, $S_p$) and the β-decay *Q* value.

The relationship between the separation energies and the β-decay Q value ($Q_β$) in a related pair of nuclei is shown in Fig. 1. The observing window of the present $^A$C(p,n)$^A$N reaction is shown by the shaded area between the two arrows pointing from the ground state of the $^A$N nucleus to its excited states below the proton separation energy. The neutron separation energy $S_n$($^A$N) for a neutron-rich N isotope is always smaller than the proton separation energy $S_p$($^A$N), so neutron evaporation may occur within this window though the final nucleus remains an N isotope. The separation energies and the $Q_β$ are related as

$$S_p(\text{N}) + 0.782 = Q_β + S_n(\text{C}), \tag{2}$$

where 0.782 MeV is the mass difference between a neutron and hydrogen ($M_n - M_H$). For a neutron-rich nucleus the neutron separation energy is small. In particular, the neutron separation energy is about 1 MeV for nuclei along the R-process. Therefore, in such nuclei the (p,n) transition window is very close to the β-decay window determined by the $Q_β$-value. Therefore, $σ_{ex}$ may be closely related to the total β-transition strength for nuclei near the R-process path.

In an effort to study the above, the isotope dependences of $σ_{ex}$ for C-isotopes for *A* = 12–19 were measured using hydrogen and carbon targets at the SIS-18/FRS facility at GSI. Incident beams of 1 GeV/nucleon $^{40}$Ar and $^{22}$Ne were used to produce secondary beams of C isotopes at around 950 MeV/nucleon. The production target was a 5 g/cm$^2$ thick

Be plate. The measurements were made at the final achromatic focus of FRS [14] after selection of C isotopes. The intensity of total secondary nuclei at the secondary target was kept below a few thousand per second at maximum. The primary beam intensity was accordingly changed for different setting of isotopes but ranged from $10^8$ to $10^9$ per synchrotron spill that was 4 second. Details of the principle and the method of nuclear separation are described in Ref. [14]. A schematic diagram of the present detector setup is shown in Fig. 2. In the figure, MUSIC indicates a multi-sampling ion chamber, TPC indicates a time-projection chamber, Sci indicates a plastic scintillation detector, and VETO is a plastic scintillation detector with a hole in the middle. Each MUSIC is segmented into 8 cells and the signal of each cell was read by an anode pad. Incident particles are identified by $\Delta E$ signals from MUSIC1 and the time-of-flight (TOF) is determined by the time difference between the signal from the first plastic scintillator (Sci1) and the signal from the plastic scintillator placed at S2, which is the dispersive focus of the FRS located 36 m upstream from Sci1. The track of an incident particle is determined by TPC1 and TPC2, which are placed before and after MUSIC1. From these measurements, the incident positions and incident angles of a particle at the target can be determined. This information is used to select incident particles that satisfy the condition determined by the following detectors. The VETO counter is used to reject events for which the incident carbons are associated with other charged particles. The $Z$ resolution ($\sigma_Z$) of MUSIC1 is 0.12 when all 8 cells of the signals are added. The number of incident nuclei is determined by selecting good incident nuclei using TOF, $\Delta E$ by MUSIC1, and the incident position and angle. The mixing of other nuclides in the incident C isotopes is always less than $10^{-4}$ and thus no effect of such contamination is expected in the measured cross sections.

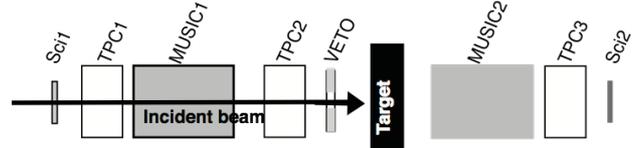

Fig. 2 Experimental setup. Sci: plastic scintillation detector, TPC: time-projection chamber, MUSIC: multi-sampling ion chamber.

The $\Delta E$ signals from MUSIC2 are used to determine the $Z$ of the particles after the reaction target located just upstream of MUSIC2. MUSIC2 measures 8 layers of $\Delta E$ for a particle. The active area of MUSIC2 is $200 \times 80$ mm$^2$ and the active length is 400 mm. The smallest covering polar angle of MUSIC2 from the target was 106 mrad, which is large enough to cover almost all the projectile fragments. The position distribution of $Z = 7$ particles was also measured by the position sensitive detector TPC3 after MUSIC2, which confirmed that all the N production events are well contained within the MUSIC2 active area. Two types of reaction targets were used: a graphite plate of 4.010 g/cm$^2$ and a polyethylene plate of 3.625 g/cm$^2$ in thicknesses. Data were also accumulated without a target (empty target) to estimate the number of reactions that occur at places other than at the target. The experimental setup is the same as that presented in previous papers. [15,16]

Figure 3 shows a $\Delta E$ spectra of the MUSIC2 detector after the C target for a measurement with an incident $^{18}$C beam. The upper panel shows a $\Delta E$ spectrum obtained by summing all the signals from the 8 layers of MUSIC2. The highest peak in the histogram is from the non-interacting $^{18}$C and a small peak at the right hand side of the $^{18}$C is the peak for $Z = 7$ nuclei. The $Z$ resolution ($\sigma_Z$) of MUSIC2 is 0.12. The number of produced $Z = 7$ nuclei is determined from the total count of events that have an energy loss larger than the energy determined by the minimum counts of the spectrum between $Z = 6$ and $Z = 7$. The $\Delta E$ signals in MUSIC2 can also be divided into the front 4 layers (M2F) and the back 4 layers (M2B). The lower panel of Fig. 3 shows the scatter plot of $\Delta E$ for M2F and M2B. Almost all of the $Z = 7$ events show a consistent $\Delta E$ between M2F and M2B. This indicates that the loss of $Z = 7$ particles, by scattering, reactions, or anything else in the detector is small. The amount of loss is estimated to be less than 5% of the events and thus is not corrected for in the estimates of the cross sections. The cross sections $\sigma_{N=7}$ for C and polyethylene targets were determined after subtracting the empty target background. The background here mainly comes from the reactions occurred in the detectors

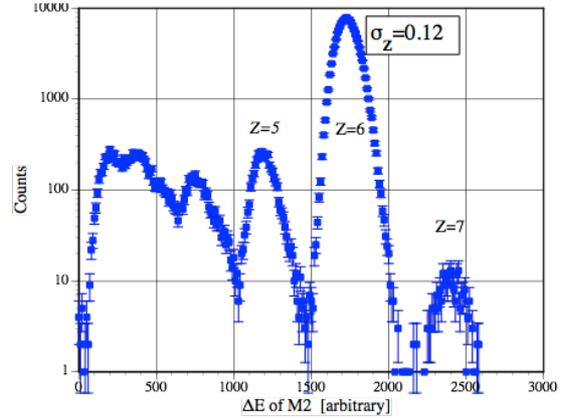
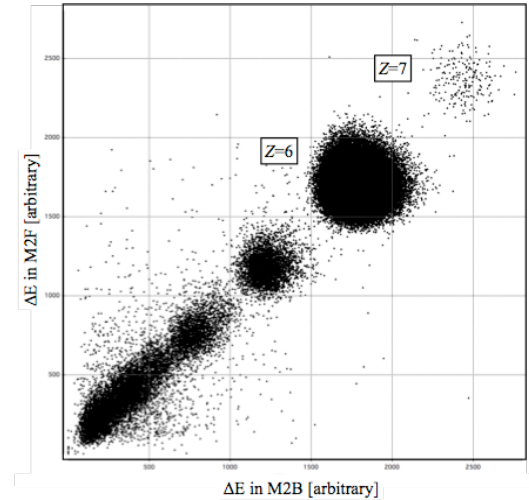

Fig. 3 $\Delta E$ spectra in MUSIC2 for $^{18}$C incident on a C target. Upper panel: pulse height spectrum of sum of all 8 layers (M2) in MUSIC2. Lower panel: scatter plot of the sum of the front 4 layers (M2F) and the sum of the back 4 layers (M2B).

**Table I Observed production cross section of N from C isotopes.**

| A | $\sigma_{z=7}$ (H) [mb] | $\sigma_{z=7}$ (C) [mb] | $S_p$ (N) [MeV] | $S_n$ (C) [MeV] | $Q_{\beta^-}$ (C) [MeV] | IAS [MeV] | $N_{tr}$ |
|---|---|---|---|---|---|---|---|
| 12 | 0.11 ± 0.07 | 0 ± 0.08 | 0.6 | 18.72 | -17.34 | - | 4 |
| 13 | 0 ± 0.2 | 0.33 ± 0.33 | 1.94 | 4.95 | -2.22 | 0 | 1 |
| 14 | 0.30 ± 0.08 | 0.36 ± 0.08 | 7.55 | 8.18 | 0.16 | 2.313 | 2 |
| 15 | 0.29 ± 0.09 | 0.61 ± 0.11 | 10.21 | 1.22 | 9.77 | 11.615 | 3 |
| 16 | 0.76 ± 0.15 | 1.91 ± 0.20 | 11.48 | 4.25 | 8.01 | 9.93 | 14 |
| 17 | 1.45 ± 0.29 | 3.14 ± 0.36 | 13.13 | 0.73 | 13.16 | unknown | 26 |
| 18 | 1.27 ± 0.21 | 5.87 ± 0.27 | 15.21 | 4.19 | 11.81 | unknown | 38 |
| 19 | 1.88 ± 0.65 | 7.66 ± 0.80 | 16.97 | 0.16 | 16.55 | unknown | 50 |

*A*: mass number of incident carbon, $\sigma_{z=7}$ (H): cross section with H target, $\sigma_{z=7}$ (C): cross section with C target, $S_p$: proton separation energy, IAS: excitation energy of isobaric analog state of $^A$C in $^A$N, $N_{tr}$: number of possible $L = 0$ transitions.

after the incident identifications and the admixture of Z=7 nuclides in the incident beam, if any. The typical rate of the background was 6 x 10$^{-5}$ of the incident beam. The cross section for a proton target was obtained by subtracting the C target cross section from that of the polyethylene target.

The determined cross sections are listed in Table I and are presented in Fig. 4. The cross section increases rapidly as the number of neutrons increases in C isotopes. The rate of increase for the proton target is faster than linear with respect to the neutron number. The cross section increases even faster for the C target.

We first consider reactions with the H target. The present measurements do not allow the mass number of N isotopes in the final states to be determined; therefore, neutron emissions followed by (p,n) reactions are not distinguished. In other words, (p,n) reactions below the proton emission threshold are all integrated in the measured cross sections whether or not they emit neutrons. At the present beam energy, the rate of proton capture reactions followed by neutron evaporation is considered to be negligibly small [11,12]. Therefore, almost all the reactions for the production of N isotopes are charge exchange (p,n) reactions. The (p,n) cross section at small scattering angles at the present high energy is expected to be dominated by Fermi and GT transitions. The present measurement, however, covers almost all the scattering angle of the (p,n) section therefore the transitions correspond to other types of selection rules may be included. In the following however we assumed that the main contributions are mainly from "allowed" transitions that is dominant in beta decays. We have started collaboration with theory to estimate the contributions of such transitions as well as GT and F transitions in the total-charge-changing cross sections [17].

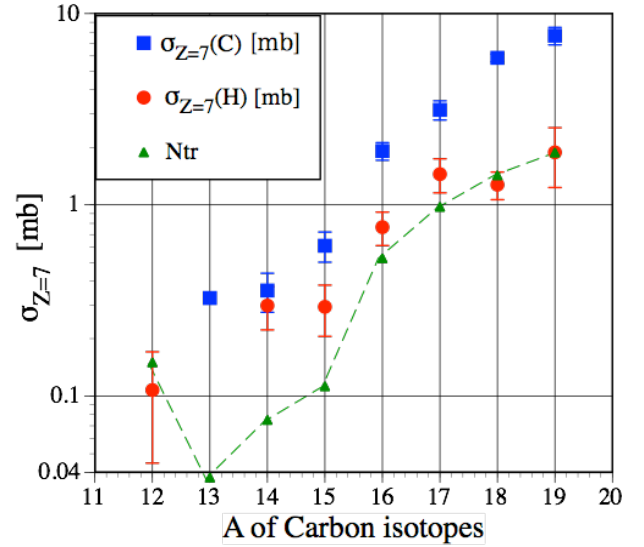

Fig. 4 Observed charge exchange cross sections of C isotopes on H and C targets. Arrows for *A*=12 and 13 indicates that the error bars extend below the bottom of the figure. Two simple model expectations of neutron number dependence of the cross sections are shown by the dashed line. (See text for an explanation.)

Table II. Beta decay strength of C isotopes

| β-transition | Final state | log*ft* | 1/*ft* [×10⁻⁷] | $\bar{\sigma}_{ex\beta}$ *8 | A for sum |
|---|---|---|---|---|---|
| ¹²N (1⁺) -> ¹²C *¹ | gs (0⁺) | 4.12 ± 0.003 | 2276 ± 16 | 1.403 ± 0.009 | 12 |
| ¹³N (1/2⁻) -> ¹³C *² | gs (1/2⁻) | 3.667 ± 0.001 | 2153 ± 5 | 0.427 ± 0.001 | 13 |
| ¹⁴C (0⁺) -> ¹⁴N *³ | gs (1⁺) | 9.04 ± ? | 0.009 ± ? | | |
| ¹⁴O (0⁺) -> ¹⁴N *³ | gs (1⁺) | 7.266 ± 0.009 | 0.524 ± 0.011 | | |
| (Mirror) | 2.31 (0⁺) | 3.4892 ± 0.0002 | 3241.9 ± 1.5 | | |
| | 3.95 (1⁺) | 3.15 ± 0.02 | 7080 ± 330 | 4.56 ± 0.14 *⁷ | 14 |
| ¹⁵C (1/2⁺) -> ¹⁵N *⁴ | gs (1/2⁻) | 5.99 ± 0.03 | 10.2 ± 0.7 | | |
| | 5.30 (1/2⁺) | 4.11 ± 0.01 | 776 ± 18 | | |
| | 7.30 (3/2⁺) | 6.89 ± 0.05 | 1.29 ± 0.15 | | |
| | 8.31 (1/2⁺) | 5.18 ± 0.05 | 66 ± 8 | | |
| | 8.57 (3/2⁺) | 5.34 ± 0.07 | 46 ± 7 | | |
| | 9.05 (1/2⁺) | 4.05 ± 0.04 | 891 ± 82 | 1.09 ± 0.05 *⁷ | 15 |
| ¹⁶C (0⁺) *⁵ | gs (2⁻) | - | | | |
| | 0.12 (0⁻) | 6.7 ± 0.07 | 2.00 ± 0.32 | | |
| | 3.35 (1⁺) | 3.551 ± 0.012 | 2812 ± 78 | | |
| | 4.32 (1⁺) | 3.83 ± 0.05 | 1480 ± 170 | 2.63 ± 0.12 *⁷ | 16 |
| ¹⁸C (0⁺) *⁶ | gs (1⁻) | - | | | |
| | 1.735 ((2⁺)) | 5.2 ± 0.4 | 63 ± 58 | | |
| | 2.614 (1⁺) | 4.08 ± 0.08 | 830 ± 150 | 0.51 ± 0.09 *⁷ | 18 |

*¹ Ref. [18], *² Ref. [19], *³ Ref. [20], *⁴ Ref. [21], *⁵ Ref. [22], *⁶ Ref. [23]
*⁷ Value obtained only from the *ft* values of ¹⁴O β decay. It is the summed values of all listed states.
*⁸ See Eqs. (4)–(10) for definitions
?: The error is not shown in reference [20].

Next we compare these cross sections with observed beta-decay transitions. The known log *ft* values are listed in Table II. The beta-decay transition strength $B(\alpha)$ in Eq. (1) is related to the *ft* value by

$$G_V^2 B(\text{F}) + G_A^2 B(\text{GT}) = \frac{K}{ft}, \qquad (3)$$

where $K$, $G_V$, and $G_A$ are the coupling constants of beta decay and are common for all nuclei. The ratios of the axial-vector coupling constant and the vector coupling constant are $R = (G_A/G_V)^2 = 1.56 \pm 0.2$ and $K/(G_V)^2 = 6163 \pm 4$ s. Because we are discussing the relationship between the summed cross sections of charge exchange and beta decays, it is in general not possible to separate Fermi and GT transitions. Under this assumption:

$$B(\text{F}) + RB(\text{GT}) = \frac{6163}{ft}. \qquad (4)$$

Because the present measurement deals with values of $\sigma_{ex}$ below the proton emission threshold ($S_p$), which is close to the corresponding beta decay Q-value ($Q_\beta$), comparisons between the integrated β strength below $S_p$ and $\sigma_{ex}$ are meaningful. Because the unit cross sections of Fermi and GT transitions are different, the total charge-exchange cross section evaluated from beta-decay $\sigma_{ex\beta}$ can be written as,

$$\sigma_{ex\beta} = \sum_{\text{all transitions}} \left[ \hat{\sigma}_F B(\text{F}) + \hat{\sigma}_{GT} RB(\text{GT}) \right]. \tag{5}$$

For a pure Fermi or GT transition, adding these transition strengths are straight forward. However, for a mixed transition such as $^{13}$C -> $^{13}$N, this addition should be treated carefully. Note that the unit cross sections $\hat{\sigma}_F$ and $\hat{\sigma}_{GT}$ are not the same, but the unit cross sections of the Fermi transition is much smaller than that of GT transitions: $\hat{\sigma}_F / \hat{\sigma}_{GT} \sim 1/10$ [4]. Therefore,

$$\sigma_{ex\beta} = \sum_{\text{all transitions}} \hat{\sigma}_{GT} \left[ B(\text{F})/10 + RB(\text{GT}) \right]. \tag{6}$$

For a pure Fermi transition, the partial strength for a transition $\sigma_{ex\beta}^k(F)$ is proportional to

$$\sigma_{ex\beta}^k(\text{F}) = \hat{\sigma}_{GT} B(\text{F})/10 = \hat{\sigma}_{GT} \frac{616.3}{ft},$$

and

$$\bar{\sigma}_{ex\beta}^k(\text{F}) \equiv \sigma_{ex\beta}^k(\text{F})/\hat{\sigma}_{GT} = \frac{616.3}{ft}, \tag{7}$$

where $k$ indicates an individual transition and $\bar{\sigma}_{ex\beta}^k(\text{F})$ is the normalized strength of a Fermi transition. The normalized strength for a pure GT transition is then

$$\bar{\sigma}_{ex\beta}^k(\text{GT}) \equiv \sigma_{ex\beta}^k(\text{GT})/\hat{\sigma}_{GT} = \frac{6163}{ft}, \tag{8}$$

and the normalized total charge exchange cross section is,

$$\bar{\sigma}_{ex\beta} = \sum_k [\bar{\sigma}_{ex\beta}^k(\text{F}) + \bar{\sigma}_{ex\beta}^k(\text{GT})]. \tag{9}$$

Note that care should be taken for a mixed transition; the partial Fermi and GT amplitudes in the transition should be calculated individually and then added.

In the following we examine such relationships for pairs of C and N nuclei:

- $^{12}$C is stable so no beta transitions to $^{12}$N can be observed. The transitions that affect the Cex reaction occur below 0.601 MeV excitation energy in $^{12}$N. Only the ground state of $^{12}$N exists within this range, so the β transition strength can be obtained from the β decay of $^{12}$N to the ground state of $^{12}$C, which is a pure GT transition. The observed log $ft$ value is shown in Table II. To obtain $B$(GT) for $^{12}$C -> $^{12}$N, a spin factor $(2J_{^{12}N}+1)/(2J_{^{12}C}+1)$ should be included because

$$B(\text{GT}) = \left| \left\langle f \left\| \sum_k \sigma_k t_k \right\| i \right\rangle \right|^2 / (2J_i + 1), \tag{10}$$

where $|i\rangle$ and $|f\rangle$ are the initial and final states, respectively, $\sum_k \sigma_k t_k$ is the GT operator, $k$ is the nucleon index, $\sigma_k$ is the Pauli spin operator, and $t_k$ is the isospin operator. The spin factor corrected $ft$ value is used to obtain values of $\bar{\sigma}_{ex\beta}$, shown in Table II.

- $^{13}$C is also stable for beta transitions. Only the ground state of $^{13}$N is below the proton emission threshold, very similar to the $^{12}$C case, and only transitions between the ground states contribute to the Cex reaction, which includes Fermi and GT mixed transitions between mirror states. In this case, $B(\text{F}) = 1$ and $B(\text{GT})$ can be calculated directly from the beta decay transition. For the present $^{13}$N and $^{13}$C case, $RB(\text{GT}) = 0.327$ from the $ft$ value listed in Table II. Therefore, the Fermi transition strength is larger than that of the GT transition. The spin factor is 1 because the spins of the initial and final states are both 1/2. The $\bar{\sigma}_{ex\beta}$ value with mixed transitions is shown in Table II.

- $^{14}$C decays to $^{14}$N, but only to the ground state, and this transition is known to be very weak. The proton emission threshold for $^{14}$N is $E_x = 7.55$ MeV and many states exist below this excitation energy. Although only beta transitions between the ground states can be observed for $^{14}$C, the mirror nucleus $^{14}$O exhibits transitions up to the $E_x = 3.95$

MeV state. The 2.313 MeV state in $^{14}$N is the IAS of the $^{14}$C and $^{14}$O ground states and thus the transition is a super-allowed Fermi transition. The transition to 3.95 MeV is also very strong: log $ft$ = 3.15. The spin factor is 1, so the sum of 1/$ft$ is very large compared with that of $^{12}$C and $^{13}$C. In addition, although β decay is not possible due to the Q-value, a $1^+$ state exists at $E_x$ = 6.024 MeV. The transition to this state is expected to contribute to the Cex transition and thus the obtained $\bar{\sigma}_{ex\beta}$ shown in Table II may be an underestimation. Although the $ft$ values of the mirror transitions may have some asymmetry of 10–20% [24,25], this is low enough that it does not affect the following discussions.

The calculated strength from the $ft$ values $\bar{\sigma}_{ex\beta}$ is much larger than that for $^{12}$C and $^{13}$C. In fact the observed $\sigma_{ex}$ increases suddenly between $^{13}$C and $^{14}$C, which is consistent with the increase of the calculated strength. Taddeucci [4] found that the unit cross sections varies for the $^{12}$C, $^{13}$C, and $^{14}$C isotopes. The difference is about a factor of 1.5 between $^{12}$C and $^{13}$C and almost 1 between $^{12}$C and $^{14}$C, and is much smaller than the present increase of the beta-transition strength of more than a factor of 5. The value of $\sigma_{ex}$ with a proton target and the strength presently obtained from β-decay $\bar{\sigma}_{ex\beta}$ are presented in Fig. 5. The strengths obtained from β-decays are normalized for $^{12}$C. The change of cross section between $^{12}$C and $^{14}$C agrees well with the change in the β-decay strength within experimental errors although the uncertainty of $^{12}$C cross section is large (~60% error). The small value of $\bar{\sigma}_{ex\beta}$ is also consistent with the cross section data. Although the uncertainties are large, the comparison indicates a similarity between β strength and $\sigma_{ex}$ within factor of two.

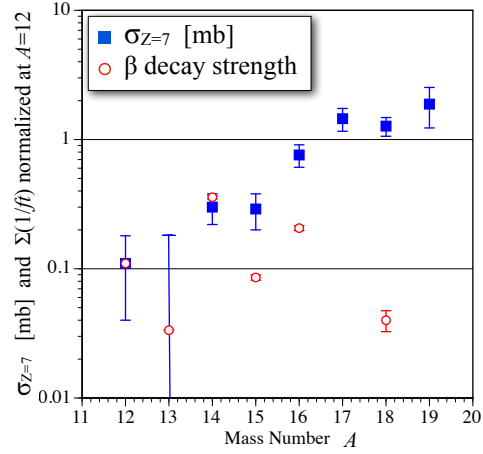

Fig. 5 Comparison between β decay strengths $\bar{\sigma}_{ex\beta}$ and $\sigma_{ex}$ of C isotopes with proton target. The central value of $A$=13 cross section is 0. The upper limit of the error of teh cross section is shown in the figure.

- $^{15}$C decay exhibits branches to several excited states of $^{15}$N. Beta decays are observed up to the 9.05 MeV ($1/2^+$) state. All of the transitions have log $ft$ values larger than 4 and thus are weak transitions. The proton separation energy is 10.21 MeV and there are several states with $1/2^+$ and $3/2^+$ spin parities. Some of these states may have larger strength; therefore, the strength determined from known β decays is expected to be smaller than the strength seen in the Cex reaction. This is consistent with the comparison in Fig. 5 although the amount of missing strength cannot be estimated. $\sigma_{ex}$ is smaller than the expected strength from an interpolation between $^{14}$C and $^{16}$C. This may be understood as follows. The IAS of $^{15}$C in $^{15}$N is located at $E_{ex}$ = 11.615 MeV and is above the proton emission threshold. It is a mixed transition of GT and F and thus is expected to have large contribution to the (p, n) cross section. The transition to this state, however, does not contribute to the present charge-changing cross section. This is in contrast to the $^{14}$C and $^{16}$C cases for which the IASs contributes to the cross sections.

The reason why the IAS contributes for $^{14}$C but not for $^{15}$C is due to the effect of staggering of the $Q_\beta$ value for even and odd neutron numbers in the isotopes. Although $Q_\beta$ shows staggering, $S_p$ shows no staggering as expected from the paring-energy term in the Bethe–Weizsäcker mass formula. The IAS is located at excitation energy slightly above a $Q_\beta$ value, so it can jump above and below the proton emission threshold. For $^{14}$C and $^{16}$C, the IASs are below the proton emission threshold and expected to be so also for $^{18}$C. In contrast they are above the threshold in $^{15}$C and expected to be so in $^{17}$C and $^{19}$C. The relation between $Q_\beta$, $S_p$, and the excitation energy of the IAS is listed in Table I and shown in Fig. 6. The even–odd change of distance between $Q_\beta$ and $S_p$ is clearly seen.

- $^{16}$C β decays are observed up to the 4.32 MeV excited state of $^{16}$N, including transitions to 3.35 and 4.32 MeV, which are GT transitions with large transition strengths. Many other $1^+$ states exist below the proton emission threshold ($S_p$ = 11.45 MeV). Much more strength is expected to contribute to $\sigma_{ex}$. The IAS is at 9.93 MeV and thus also contributes to some extent to the cross section. The contribution from IAS may be smaller than that for $^{15}$C because it is pure Fermi transition. The observed large increase of $\sigma_{ex}$ from that of $^{15}$C could thus be due to missing GT transition contributions.

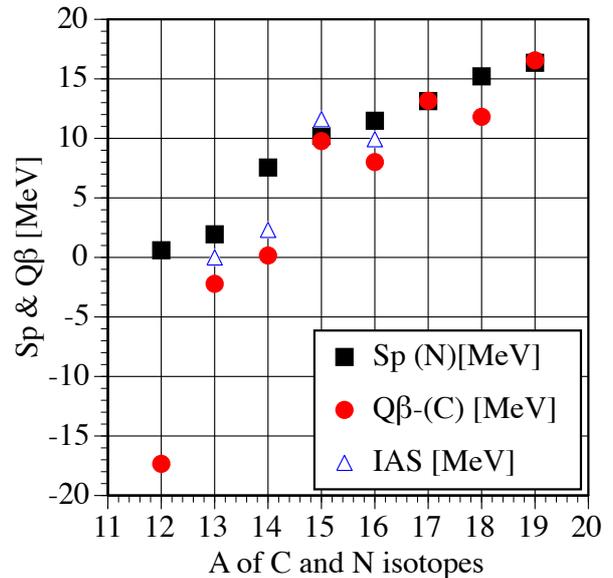

Fig. 6 $Q_\beta$ of C isotopes and $S_p$ of N isotopes.

- $^{17}$C beta decay branches have been observed but only the branching ratio among γ-decaying states and the total branching ratio to neutron-emission channels have been determined but the log $ft$ values have not yet been determined. [26]
- $^{18}$C beta decay log $ft$ values have been reported only for two transitions, to the 1.735 and 2.614 MeV states. Many transitions go to neutron emitting states and the branching ratio to this channel is not convincing. The known log $ft$ values are listed in Table II but it is obvious that large strength is missing.

To summarize the comparison between $\sigma_{ex}$ and $\bar{\sigma}_{ex\beta}$, they behave similarly to each other for $A$ = 12, 13, 14 where most of the beta strength below the proton emission threshold is known. This suggests a reasonable proportionality between $\sigma_{ex}$ and $\bar{\sigma}_{ex\beta}$. For $A > 14$ it is obvious that the β strengths that contribute to $\sigma_{ex}$ could not be obtained by β-decay measurements. The calculated $\bar{\sigma}_{ex\beta}$ values are thus much smaller than $\sigma_{ex}$ in those cases. If we believe that $\sigma_{ex}$ is a good reflection of the integrated β strength, supported somewhat by the cases of $A$ = 12, 13,14, the observed $\sigma_{ex}$ values for heavier C isotopes represent the integrated β strength. As already mentioned the uncertainties for $^{12}$C, and $^{13}$C are large, therefore this assumption has to be examined more carefully in the future.

The proton separation energies of $^{12,13}$N are small, as seen in Table I. The transition windows in those nuclei are thus very narrow and the observed cross sections are much smaller than the sum rule values. For example, in $^{13}$C, a neutron in a $p_{1/2}$ orbital can be transferred to a proton in the same $p_{1/2}$ orbital in a single particle model to form $^{13}$N. However a transition of a neutron from a $p_{3/2}$ to a proton in a $p_{1/2}$ orbital cannot be observed in the present measurements because the particle–hole excitation energy is about 15 MeV (the mass difference between $^{12}$C and $^{12}$N) and thus this state decays by proton emission. For $^{15}$C ($J^\pi = 1/2^+$), the last neutron orbital is considered to be $s_{1/2}$ in the simplest model, and new transitions to the $s_{1/2}$ orbital are added to the transition between the neutron and proton $p_{1/2}$ states. After $^{16}$C, additional neutrons are considered to be in the sd shell and all the sd-shell orbitals are considered to contribute. Hence we consider that the cross section may be proportional to the number of possible transitions to sd shell orbitals. In this way we can extrapolate the strength for nuclei in which β decay observations are restricted. This counting model, again normalized for $^{19}$C, is shown by the dashed curve in Fig. 4. It has a much stronger $N$ dependence than the observed $\sigma_{ex}$. More detailed model calculations using realistic wave functions and reaction mechanisms are therefore required to explain the H target cross sections.

A peculiar observation is a much stronger $N$ dependence of $\sigma_{ex}$ with a C target. As shown in Fig. 7, the ratio of the cross section between C and proton targets is almost 1 for $^{14}$C but more than 4 for $^{18}$C and $^{19}$C. The ratio increases almost linearly with the neutron number. N isotopes are produced by transferring a proton from a target or by charge exchange. The number of protons in a C target is 6 and thus there is a greater possibility of reaction than for a H target. If the effective number of protons in the C target is considered, the cross section may increase but this increase does not depend on the number of neutrons in the projectile. Thus, a naive consideration predicts the ratio to be constant for all isotopes.

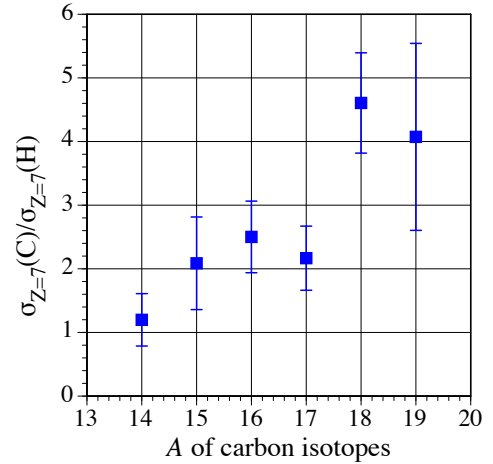

Fig. 7 Ratio of the cross sections between C and H targets. A faster increase in C target cross section is seen as the number of neutrons in the projectiles increases.

A possible difference between a proton target and a C target is the distortion. To observe the p-n exchange reaction without losing protons from the projectile, no additional collisions between other nucleons should occur simultaneously. In proton–nucleus collisions, the mean free path of the proton is long and so most of the neutrons in a projectile have the same probability to hit a target proton. Therefore the p–n scattering probability should be proportional to the number of neutrons. In contrast, a C target has many nucleons so that additional nucleon collisions occur frequently. Thus, single p–n collisions occur only at the surface (or at large impact parameters). The neutron-number dependence of the single p–n collision cross section may, therefore, be different for proton and nuclear targets.

In the following, we consider how the distortion affects the neutron number dependence of the $C_{ex}$ cross section. We use a Glauber model to evaluate the distortion [27]. Under the Glauber model the reaction cross section ($\sigma_R$) of nucleus–nucleus collisions is calculated by the formula

$$\sigma_R = \int [1 - T(\mathbf{b})] d\mathbf{b} , \qquad (11)$$

where the transmission function $T(\mathbf{b})$ is the probability of no collision of any combination of nucleons in a projectile and a target for the impact parameter $\mathbf{b}$ and expressed as

$$T(\mathbf{b}) = \exp\left[-\iint \left(\sigma_{pp}\rho_{Tp}(\mathbf{r})\rho_{Pp}(\mathbf{r}-\mathbf{b}) + \sigma_{pn}\rho_{Tp}(\mathbf{r})\rho_{Pn}(\mathbf{r}-\mathbf{b}) + \sigma_{np}\rho_{Tn}(\mathbf{r})\rho_{Pp}(\mathbf{r}-\mathbf{b}) + \sigma_{nn}\rho_{Tn}(\mathbf{r})\rho_{Pn}(\mathbf{r}-\mathbf{b})\right) d\mathbf{r}\right] \quad (12)$$

where $\sigma_{ij}$ are free nucleon–nucleon cross sections and $\rho_{Ti}$ and $\rho_{Pi}$ are the z-axis-integrated density of an $i$-nucleon (p or n) in the target and the projectile, respectively, where $z$ is the direction of the incident beam. $[1-T(\mathbf{b})]$ is then the total distortion function.

For neutrons in a projectile and protons in a target, the probability of collision is calculated from the density overlap as $\iint \sigma_{pn}\rho_{Tp}(\mathbf{r})\rho_{Pn}(\mathbf{r}-\mathbf{b})d\mathbf{r}$. The average number ($\lambda$) of the product of the numbers of protons ($n_p$) and neutrons ($n_n$) in a collision ($n_p n_n$) is calculated by dividing the probability by $\sigma_{pn}$, as

$$\lambda \equiv <n_p n_n> = \iint \rho_{Tp}(\mathbf{r})\rho_{Pn}(\mathbf{r}-\mathbf{b})d\mathbf{r} \quad (13)$$

The probability for scattering only one proton and one neutron (1on1) within this collision is calculated by the Poisson distribution,

$$P_\lambda(\kappa) = \frac{\lambda^k}{k!} e^{-\lambda}, \text{ with } k=1. \quad (14)$$

The probability $\mathcal{P}^1(\mathbf{b})$ of a 1on1 collision in a reaction as a function of an impact parameter can then be calculated as

$$\mathcal{P}^1(b) = \lambda e^{-\lambda}. \quad (15)$$

The distortion for the reaction is calculated using $T(\mathbf{b})$ modified for the present reaction. The effect of a 1on1 collision should be subtracted, the second term of $T(\mathbf{b})$ in Eq.(9), from for the distortion. Also the neutron-neutron collision term, the fourth term, has to be removed because we do not observe the removal of neutrons. Therefore the transmission for 1on1 collision $T_{np}(\mathbf{b})$ is written as,

$$T_{np}(\mathbf{b}) = \exp\left[-\iint \left(\sigma_{pp}\rho_{Tp}(\mathbf{r})\rho_{Pp}(\mathbf{r}-\mathbf{b}) + \sigma_{np}\rho_{Tn}(\mathbf{r})\rho_{Pp}(\mathbf{r}-\mathbf{b})\right) d\mathbf{r}\right], \quad (16)$$

where terms of collisions of projectile proton are contributing.

The probability of one neutron in the projectile colliding with a proton in the target without any other nucleon collision is calculated by $\mathcal{P}^1(\mathbf{b})T(\mathbf{b})$. The cross section ($\sigma_{ex,G}$) of a charge exchange reaction (p,n) is then proportional to

$$\sigma_{ex,G} = 2\pi \int \mathcal{P}^1(b) T_{-np}(b) b\, db. \quad (17)$$

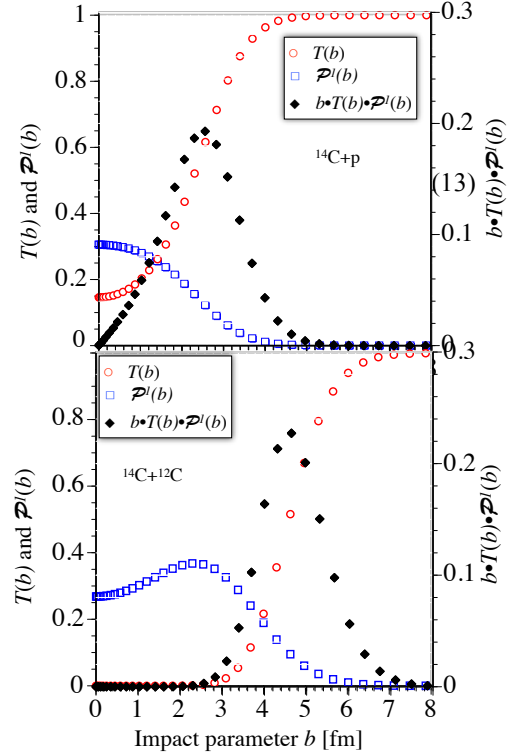

Fig. 8 Impact parameter dependence of $\mathcal{P}^1(b)$ and $T(b)$ and $b*\mathcal{P}^1(b)*T(b)$. The upper panel presents the result for $^{14}C+p$ and the lower panel presents the result for $^{14}C+^{12}C$. See text for definitions.

Such calculations have been made for C and H targets. The density distribution of a proton is assumed to be Gaussian with a root-mean-square radius of 0.8 fm.

Figure 8 shows the results of $\mathcal{P}^1(b)$, $T(b)$ and $b*\mathcal{P}^1(b)*T(b)$. For a proton target (upper panel in Fig. 8) $\mathcal{P}^1(b)$ has a maximum at zero impact parameter, reflecting a long proton mean free path. However $\mathcal{P}^1(b)*T(b)$ shows a dominance at the surface of the nucleus due to $T(b)$, although some contribution still remains from $b=0$. For $^{14}C+^{12}C$ collisions (lower panel in Fig. 8), 1on1 p–n collisions occur only in a narrow region of a large impact parameter where the surfaces of two nuclei are just touching. Although there are slight differences in $b*\mathcal{P}^1(b)*T(b)$, the surface character of the collision is essentially the same for both H and C targets.

$\sigma_{ex,G}$ is calculated under two different density distributions using a harmonic oscillator model. For the first model, the density distributions of protons and neutrons are described by the same size parameter. The size parameters for C-isotopes were determined by analyzing the interaction cross sections [28, 29]. Differences in the proton and neutron distributions arise only from the difference in the $N$ and $Z$ values. The obtained values of $\sigma_{ex,G}$ under this model are plotted in Fig. 9 labeled as $\sigma^1_{ex,G}$. It should be noted that these are not the absolute values of the cross section but the relative values. The values of $\sigma^1_{ex,G}$ in both the C+p and C+C cases smoothly increases as the neutron number increase. The lower panel of Fig. 9 shows the ratio of C+C to C+p that should be compared with the cross section ratio in Fig. 7. The calculated ratio is almost the same for all isotopes in contrast with the measured ratio of the cross sections, which increases by a factor of 4 in $^{19}C$. Therefore, the distortion is not a likely origin of the stronger n-dependence of the N isotope production with a C target. We also tested the density model dependence of the distortion effect using more realistic density distributions of proton and neutrons in C isotopes using the recently measured charge-changing cross sections [30] and the interaction cross sections. Thicker neutron skins for $^{16-18}$C and a neutron halo in $^{19}$C are reflected in the oscillator size of neutrons. The calculated values of $\sigma^2_{ex,G}$ are also plotted in Fig. 9. It shows that the ratio of the C+C to C+p cross sections increase as mass number increases showing the importance of the effect of extra neutrons on the surface (skins and a halo). However the increase of the ratio is still much smaller than the observed increase.

Another possible reason is that other reactions are associated with clusters of nucleons in the target. For example ($^3$He, t) type reactions may transfer a proton into a projectile. However, the cross section for free $^3$He is much smaller than that of (p,n) reactions. The (d,2n) reaction was also observed in a $^{14}$Be incident beam though the cross section was found to be very small compared with that of the (p,n) reaction [8]. Moreover, those reactions are not expected to have strong projectile neutron-number dependences. A possible stronger dependence may be considered if a reaction occurs with a 2n or more neutron-number cluster in a projectile. However, no information is available on such reactions at the present energy. As a conclusion, we do not understand the observed strong neutron-number dependence of charge exchange reactions with C targets.

In summary, we have measured the production of N from 950 MeV per nucleon C isotopes ($^{12}$C–$^{19}$C) on H and C targets for the first time. The production cross section near the projectile velocity was determined. The cross sections with an H target increase with the number of neutrons (N) in the projectile. The change of the cross section from $^{12}$C to $^{13}$C and $^{14}$C show a consistent behavior with the observed β-decay strength. The neutron-number dependence on a proton target can be roughly explained by a simple model counting the number of related single-particle orbitals and expected to be better explained with a detailed structure model of nuclei. The cross section with a C target increases much faster than for a H target. The effect of distortion due to a large mass target was found to not be sufficient to explain the difference of the neutron dependence between proton and C targets. Therefore, the present data suggests some unknown processes of production of N isotopes from high-energy C isotopes. Although the present precision of the charge-changing cross section for $^{12}$C and $^{13}$C is not good but we could see the similar behavior between the beta-decay strength and the charge changing cross sections. Taking advantage of the Q-value window in neutron rich nuclei, measurements of charge-exchange cross sections are expected to be a handy method to study the integrated β transition strength in neutron-rich nuclei.

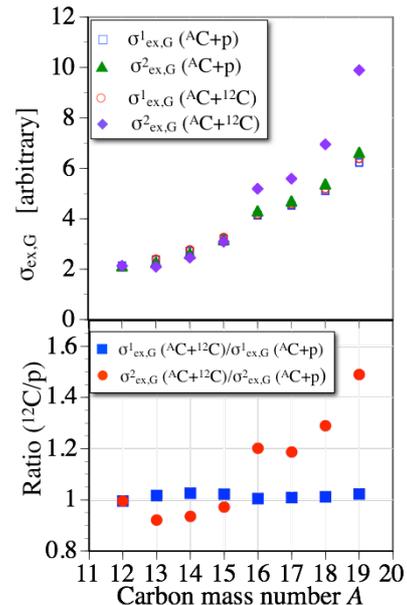

Fig. 9 $\sigma_{ex,G}$ for two different density models for $^{14}$C+p and $^{14}$C+C collisions (top panel). Ratios of the cross sections between C and p target are shown in the bottom panel. Those should be compared with the ratio presented in the Fig.7.

## Acknowledgements

The support of the PR China government and Beihang University under the Thousand Talent program is gratefully acknowledged. The experiment is partly supported by the grant-in-aid program of the Japanese government under the contract number 23224008. This project is supported by NSERC, Canada.